\documentclass[11pt, 3p, authoryear, review]{article}

\usepackage{graphicx}
\usepackage{amssymb}

\usepackage{amsmath}       
\usepackage{color}         
\usepackage{hyperref}      
\usepackage{booktabs}

\usepackage[utf8x]{inputenc}

\newcommand{\R}{\mathbb{R}}         

\newcommand{\N}{\mathbb{N}}                          
\newcommand{\Z}{\mathbb{Z}}                          
\newcommand{\set}[1]{\lbrace#1\rbrace}               

\title{A prediction interval for a function-valued forecast model}

\author{Anestis Antoniadis \and 
        Xavier Brossat     \and
        Jairo Cugliari     \and
        Jean-Michel Poggi}

\begin{document}

\maketitle

\begin{abstract}
Starting from the information contained in the shape of the load curves, we have proposed a flexible nonparametric function-valued forecast model called KWF (\textit{Kernel+Wavelet+Functional}) well suited to handle nonstationary series. The predictor can be seen as a weighted average of futures of past situations, where the weights increase with the similarity between the past situations and the actual one. In addition, this strategy provides with a simultaneous multiple horizon prediction. These weights induce a probability distribution that can be used to produce bootstrap pseudo predictions. Prediction intervals are constructed after obtaining the corresponding bootstrap pseudo prediction residuals. We develop two propositions following directly the KWF strategy and compare it to two alternative ways coming from proposals of econometricians. They construct simultaneous prediction intervals using multiple comparison corrections through the control of the family wise error (FWE) or the false discovery rate. Alternatively, such prediction intervals can be constructed bootstrapping joint probability regions. In this work we propose to obtain prediction intervals for the KWF model that are simultaneously valid for the $H$ prediction horizons that corresponds with the corresponding path forecast, making a connection between functional time series and the econometricians' framework. 
\end{abstract}

\section{Introduction}      


In the recent literature about short-term electricity demand forecast, various methods are proposed following different types of approach: time series analysis, machine learning, regression or similarity search.  Restricting our attention to 
references involving the French electricity consumption, let us cite some papers.
\cite{taylor2010triple} which uses an exponential smoothing taking into account the structure of seasonality, \cite{dordonnat2008}, \cite{dordonnat2012}
which propose a state space model that allows to track changes in the relationship between exogenous factors (mainly temperature) and the demand for electricity. In \cite{roy2005linear}, this dependence is modeled by a nonlinear regression on the temperature depending on the month, day of week and time of day. A nonparametric version  of this strategy was recently proposed in \cite{pierrot2011short} and a Bayesian approach is provided by \cite{launey2012construction}. Among the machine learning methods \cite{devaine2011expert} propose a mixture of online predictors to get forecasts adapting to  nonstationarity. The last group of models, based on similarity search is an alternative to model the dependence structure of seasonal cycles. The basic idea is that similar cases in the past have similar future consequences. For example in \cite{poggi1994prevision} the trajectory of the electricity consumption is divided into blocks of one day size. Then, using the dissimilarity measures, the blocks similar to the last observed block are searched in the past and a weight vector is built. Finally, the forecast of the next two days is obtained by a weighted average of the most similar future days where the weights are given by the weight vector. From the statistical point of view, the model is an estimate of the regression function using the kernel method, of last block against all the blocks in the past. \cite{antoniadis2006functional} extend this model to the case of stationary functional random variables. But in the context of the French electrical power demand, the hypothesis of stationarity may fail: an evolving mean level and the existence of groups that may be seen as classes of stationarity are to be considered. We explore some corrections to take into account these two main nonstationary features. Let us be a little bit more precise.

Electricity load experts naturally look at daily demand data as time functions called load curves.
In a recent paper, \cite{shang2013} uses a functional time series approach
for forecasting short-term electricity demand. This paper is
illustrated by the half-hourly electricity demand from Monday to
Sunday in South Australia. The strategy is also to consider a seasonal
univariate time series as a time series of curves, then to reduce the
dimensionality of curves by applying a functional principal component
analysis and finally, following \cite{shang2011}, the principal
component scores are forecasted using a univariate ARIMA models. In
addition, since data points in the daily electricity demand are
sequentially observed, a forecast updating method based on
nonparametric bootstrap approach is proposed to improve the accuracy
of point forecasts. 
With respect to this strategy, the scheme we propose
handles the forecasting problem in a functional way avoiding the hour by
hour processing and considers a more flexible way to construct the
distribution leading to the confidence interval for prediction.

The shape of the curves exhibits rich information about the calendar day type, 
the meteorological conditions or the existence of special electricity tariffs. 
Using the information contained in the shape of the load curves, we proposed in
\cite{antoniadis2012prevision} a flexible nonparametric function-valued forecast 
model called KWF (\textit{Kernel + Wavelet + Functional}) well suited to handle 
nonstationary series. The predictor can be seen as a weighted average of futures 
of past situations, where the weights increase with the similarity between the past
situations and the actual one. In addition, this strategy provides with a 
simultaneous multiple horizon prediction. 

Moreover, the weights from the KWF model induce a probability distribution that can be used to produce bootstrap pseudo predictions. In \cite{antoniadis2014prevision} prediction intervals are constructed after obtaining the corresponding bootstrap pseudo prediction residuals. Applied in the electrical context, the obtained intervals are not completely satisfactory. First, 
only pointwise coverage is warrantied by theoretical results. Second, 
the dependency structure of the curves is (almost) not used. 

Interestingly, econometricians have worked on a similar framework. Let $(y_t)_{t\in\mathbb{Z}}$ be a time series observed on $t = 1, \ldots, T$. Just after observing $y_T$ we want to produce a path forecast, i.e. construct a predictor of the future $H$ values of the series $ \mathbf{y}_H = (y_{t + 1}, \ldots, y_{t + H})'$ and a simultaneous prediction interval (PI) for the path, i.e. to construct a set $A\subset\mathbb{R}^H$ such that $P( \mathbf{y}_H \in A ) \geq 1 - \alpha$, for some small $\alpha \in [0, 1]$. 

The construction of simultaneous PI (i.e. intervals for a random variable) follows the guidelines of the construction of a simultaneous confidence intervals (i.e. intervals for a parameter). They can be constructed marginally on each prediction horizon or  simultaneously using multiple comparison corrections through the control of the family-wise error (FWE), using for example the Bonferroni correction, or through the control of the false discovery rate \cite{Benjamini}.

This subject has been recently studied by econometricians interested in path forecast where bootstrap pseudo prediction can be produced. \cite{staszewska2007} propose an heuristic method to eliminate the bootstrap trajectories that are extremes and then constructs the PI as the convex hull of the remaining trajectories. The PI can also be constructed by estimating a joint probability region under assumptions that can be quite strong. \cite{Jorda2010} construct this region by means of an asymptotic normal approximation. Instead, \cite{Wolf2013} construct the joint probability region
using a bootstrap strategy where the calibration is done by controlling the multiple comparison by means of a generalized notion of FWE ($k$-FWE). In this way all but a small number $k$ of horizons are warrantied to be covered. Recently,  \cite{Delattre2013} proved a theoretical result about the $k$-FWE validity.
Some more details of the mentioned methods are given in Section \ref{sec:others}.

In this work we propose to obtain PI for the KWF model that are simultaneously valid for $y_{t + 1}, \ldots, y_{t + H}$, the $H$ prediction horizons that corresponds with the corresponding path forecast. With this the connection between functional time series and the econometricians' framework is shown. 

Two references are of interest.
\cite{petiau2009} 
presents a method to obtain confidence intervals for
load forecast. It is based on the calculation of empirical quantiles
of the distribution of the relative forecast error observed in the
past. An a priori classification between days for which the load
forecast is difficult and those for which it is easier is used
together with the hour within the day of the forecast are used to
consider the past error forecasts included in the distribution
calculation. The scheme is applied to the whole
French electrical network or for each of the seven French regional
networks. With respect to this strategy, the scheme we propose
handling the forecasting problem in a functional way avoid the hour by
hour processing and it considers a more flexible way to construct the
distribution leading to the confidence interval for prediction.

Finally a related work can be mentioned: \cite{azais2010}
consider the problem of predicting the whole annual load curve of customers from easily available explanatory variables. Simultaneous confidence bands for this prediction is obtained using results on the maximum of Gaussian sequences. Here the problem is the prediction of a customer profile and not the time series forecasting.

Our paper is organized as follows. Section 2 first describes the data
and the characteristics of electricity load consumption. Then, it
recalls KWF method and describes the use KWF for bootstrap generation.
Finally, the different proposals for prediction interval construction
are introduced. Section 3 describes the main results obtained through
the numerical experiences we performed for the construction of
prediction intervals for the forecast of the French load curve.
Finally, the construction of a confidence tube is sketched in Section 4.

\section{Materials and methods}   

\subsection{Data}

Let us briefly recall some characteristics of the electricity consumption using a
French electrical dataset. 

\begin{figure}
	\includegraphics[width=\columnwidth]{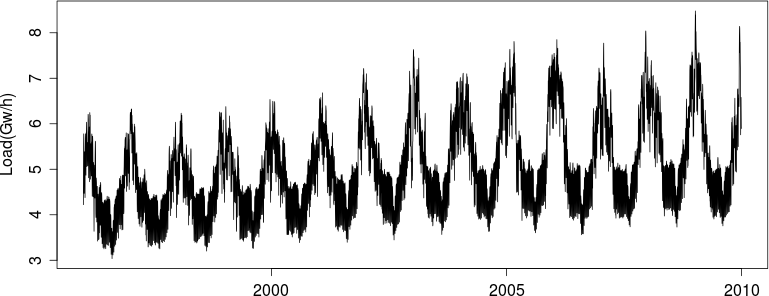}
	\caption{French electricity demand from 1996 to 2010} 
	\label{fig:longtermload}
\end{figure}

In Figure \ref{fig:longtermload} we can observe the long term evolution of the national electricity demand from 1996 to 2010. We note an increasing trend, almost linear. The annual cycle is also clearly marked with the higher levels of electricity consumption during the winter coming from the strong dependence of the consumption of electricity to weather conditions and the seasonality
of the industrial activity. 

\begin{figure}
	\includegraphics[width=\columnwidth]{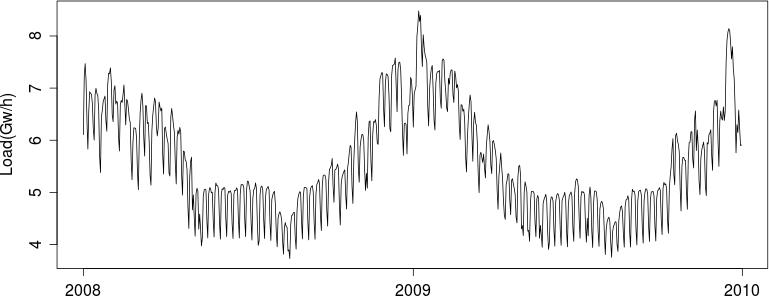}
	\caption{Two years of French electricity demand (2008 and 2009).} \label{fig:2yearsload}
\end{figure}

Zooming in on two years, we can easily distinguish a weekly cycle (Figure \ref{fig:2yearsload}). 
The economic profile of working days and weekends is reproduced by the demand with a strong increase during working days. There are other artifacts of socio-economic activity. 
As an example, during the summer, we observe two weeks during which the electricity demand is extremely low, corresponding to the summer holidays. Note 
also that the profile of the electricity demand in winter is more complex due to a large variation in demand. The impact on the forecast is a marked deterioration in performance forecasting during winter. Unfortunately, this happens during the period when the prediction errors have higher costs for electricity suppliers. 

\begin{figure}
	\includegraphics[width=\columnwidth]{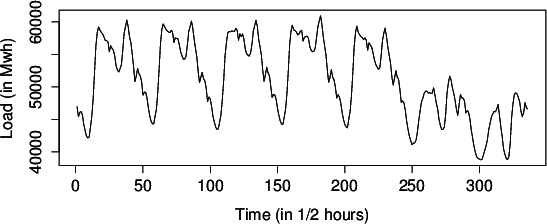}
	\caption{One typical week of load consumption in winter} \label{fig:oneweekload}
\end{figure}

\begin{figure}
	\includegraphics[width=\columnwidth]{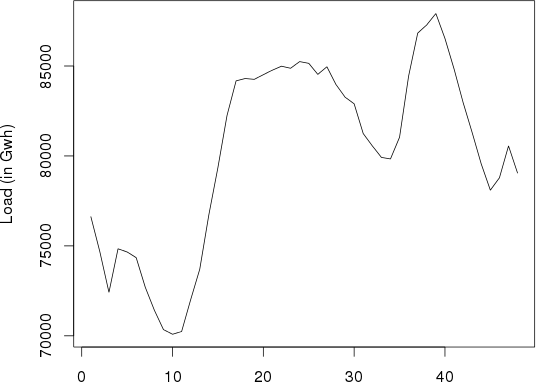}
	\caption{One typical day of load consumption in winter} \label{fig:onedayload}
\end{figure}

Figure \ref{fig:oneweekload} displays a weekly curve, and it is very easy to distinguish working days from Saturday and Sunday while Figure \ref{fig:onedayload} shows a daily curve.
Note that even at this temporal resolution, we can identify patterns: electricity consumption is lower at night, it increases between 5:00 and 9:00 am, and has a peak in the late afternoon, etc.. these 
features are identified on each daily curve. Indeed, the days for which 
it is difficult to have a good prediction of the consumption / demand for electricity are those with atypical or rare characteristics. In general, these days are also among those which cost more in terms of the prediction errors.
The power consumption depends on the climatic conditions, mainly through the dependence to the temperature. Even if in the French case, it is known to be highly sensitive to temperature and this dimension is partially taken into account in this paper but without the introduction of an exogenous variable which would greatly complicate the analysis. We consider for the numerical experiments the electricity consumption from 1996 to 2010 half-hour sampled. These data were obtained from EDF and are not publicly available. 

\subsection{KWF for bootstrap generation} \label{sec:kwf}

\subsubsection{KWF in short}

We can consider a discrete-time series as regularly sampled version of slices of a continuous process, leading to functional time series, \cite{bosq1991modelization}. 
Suppose one observes a square integrable continuous-time stochastic process $X=(X(t), t\in\R)$ over the interval $[0,T]$, $T>0$ and want to predict $X$ all over the segment $[T, T+\delta], \delta>0$. 
The prediction problem can be rephrased in terms of the function-valued discrete-time stochastic process $ Z = (Z_k, k\in\N) $, where $ \mathbb{N} = \set{ 1,2,\ldots } $, defined by 
$ Z_k(t) = X(t + (k-1)\delta) $, $k\in\N, \forall t \in [0,\delta) $ as follows. Let $(Z_k, k\in \Z)$ be a stationary sequence of random variables taking value on an separable Hilbert space. Given $Z_1, \ldots, Z_n$ 
we want to predict the future value of $Z_{n+1}$. A predictor of $Z_{n+1}$ using $Z_1, Z_2, \ldots, Z_n$ is
$    \widetilde{ Z_{n+1} }  =  \mathbb{E}[Z_{n+1} | Z_n,Z_{n-1},\ldots,Z_1] $.
The flexible framework of Autoregressive Hilbertian process of order 1 is useful. More precisely, an \textsc{arh}(1) centered process is such that at each $k$,
$Z_k = {\rho} (Z_{k-1}) + \epsilon_k$ , where $\rho$ is a compact linear operator and 
$\lbrace\epsilon_k\rbrace$  an $H-$valued strong white noise. 
Under mild conditions (\cite{bosq1991modelization}), the previous equation has a unique solution which is a strictly stationary process with innovation   $\lbrace\epsilon_k\rbrace_{k\in\Z}$ . When $Z$ is a zero-mean \textsc{arh}(1) process, the best predictor of $Z_{n+1}$ given $\{Z_1, \ldots, Z_{n-1}\}$ is:
   $ \widetilde{Z_{n+1}} = \rho(Z_n).$

Alternatively, a two step prediction algorithm can be defined to estimate $\rho(Z_n)$ using the kernel method. The step 1 is the search in the past for segments that are similar to the last observed one. One of the key points is that the dissimilarity between segments is not defined in the original domain but in the wavelet domain. Indeed, wavelets allow to cope with functional data
by hierarchically decomposing finite energy signals in a broad trend (the smooth part giving information about the mean level of the segment) plus a set of localized changes kept in the details parts giving information about the shape of the segment. Then the similarity between two observed series of length $2^J$ say $Z_m$ and $Z_l$ is obtaining by aggregating and weighting properly over the selected scales the distances between coefficients at each scale
$D( Z_m, Z_l )  =  \sum_{j=j_0}^{J-1}  2^{-j/2}  \hbox{dist}_j ( Z_m , Z_l )$,
where $\hbox{dist}_j$ is a distance associated to the scale $j$.

The second step prediction algorithm is close to the traditional kernel regression since the prediction of the approximation and detail wavelet coefficients $ \Xi_{n+1}=\{ c_{J,k}^{(n+1)}, d_{j,k}^{(n+1)} : k= 0,1,\ldots,2^j-1\}$ for $Z_{n+1}$  is obtained using the kernel $K$ and the window $h_n$ by
$ \widehat{\Xi_{n+1}} = \sum_{m=1}^{n-1} w_{m, n} \Xi_{m+1}  $
where the weights are given by
  $ w_{m,n} = \frac{    K \left( \frac{D(Z_n, Z_m)}{h_n} \right)}{
        \sum_{m=1}^{n-1} K \left( \frac{D(Z_n, Z_m)}{h_n} \right) }  $. 
Finally, the prediction of $Z_{n+1}$ is obtained through the inverse wavelet transform of the two components of the prediction: $\widehat{ \mathcal{S}_{n+1}(t)}$ the approximation part and $\widehat{ \mathcal{D}_{n+1}(t)}$ the details leading to: 
$\widehat{ {Z}_{n+1}(t)} = \widehat{ \mathcal{S}_{n+1}(t)} +
                                   \widehat{ \mathcal{D}_{n+1}(t)}$.
This strategy leads to one-segment ahead prediction and then automatically provides a simultaneous multiple horizon prediction. 

\subsubsection{Corrections to handle nonstationarity}
\begin{figure}
\centering
\includegraphics[width=\columnwidth]{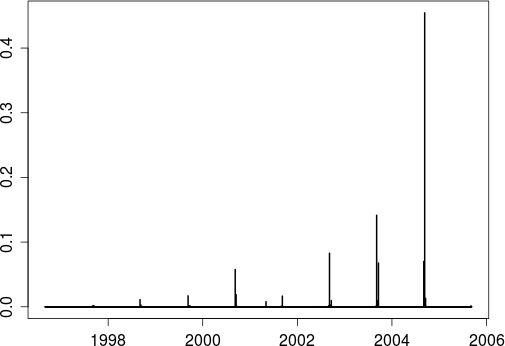}
\caption{Weight vector for the prediction of the 10th September 2010.}
\label{fig:weigths}
\end{figure}

The previous scheme is theoretically consistent and practically useful for stationary functional time series but fails when this hypothesis is not acceptable. To cope with nonstationarities, two corrections are of interest. The first one deals with the mean level, captured by the smooth part of the series, and the correction consists of predicting the one-segment increments in the mean instead of the raw value. In other words, it suffices to use $ \widehat{ \mathcal{S}_{n+1}(t)} =  \mathcal{S}_{n}(t) + 
            \sum_{m=2}^{n-1} w_{m,n} \Delta(\mathcal{S}_{m})(t) $ instead  of using 
    $ \widehat{ \mathcal{S}_{n+1}(t)} =  \sum_{m=1}^{n-1} w_{m,n} \mathcal{S}_{m+1}(t)$.
The second one is to restore stationarity by considering groups coming from calendar transitions (especially meaningful when the segment is a day), 
 or given by a clustering analysis or obtained by crossing deterministic transitions with clustering groups 
 (e.g. calendar-temperature transitions). The change is then a simple selection: after the computation of the weights $w_{m,n}$, we define the new weights
    $\tilde w_{m,n} =  w_{m,n}$  if $gr(m) = gr(n)$ and $0$ otherwise, 
    where $gr(n)$ is the group of the $n$-th segment.
Then renormalize since the weights $\tilde w_{m,n}$ must sum to 1.
These weights are highly informative namely because of the successful extraction of the information carried out on the shape of the load curves. We can illustrate this fact using Figure \ref{fig:weigths} where we plot the vector of weights for  the prediction of the Saturday 10th September 2006. The resulting positive entries of the weight vector occur on the nearby of the 10th September of the precedent years with a clearly decreasing trend for the furthest years. 
Using this fact, 
it is then possible to generate bootstrap replicates of the prediction.

\subsubsection{Bootstrap generation} \label{sec:bootKWF}

Following \cite{antoniadis2006functional}, the bootstrap pseudo-realisations which are the inputs for the construction of a prediction interval in the stationary case is given by the following steps:

\begin{enumerate}
\item Obtain the pointwise forecast $\widehat{Z}_{n+1}$ 
\item Generate $B$ bootstrap pseudo-realisations  $Z_{n + 1}^{(b)}$ 
      according to the distribution induced by the weights $\tilde w_{m,n}$, that is such that \\ 
      $P (Z_{n + 1}^{(b)} = Z_{m + 1} | Z_n) = w_{m, n}$
\end{enumerate}

Then in the nonstationary case, the main modification is to incorporate the 
approximation parts of the segment, so we continue to bootstrap the $Z$'s but we consider their 
decompositions in terms of $\mathcal{S}$'s and $\mathcal{D}$'s.

\begin{enumerate}
\item Obtain the pointwise forecast \\
$\widehat{Z}_{n+1} = \widehat{\mathcal{S}}_{n+1} + 
                       \widehat{\mathcal{D}}_{n+1} $ 
\item Generate $B$ bootstrap pseudo-realisations $Z_{n + 1}^{(b)}$ 
      such that \\
$P (Z_{n + 1}^{(b)} = Z_{m + 1} | Z_n) = \widetilde{w}_{m, n}$                     
\end{enumerate}

\begin{figure}
  \centering
  \includegraphics[width=\columnwidth]{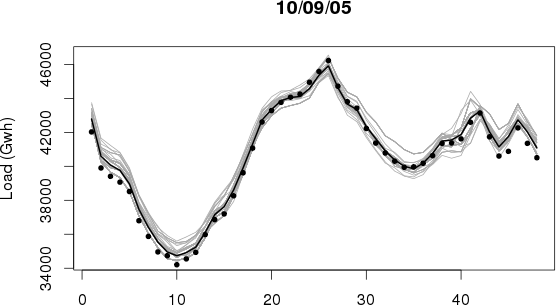}
  \caption{Pseudo realizations (gray lines) corresponding to the prediction (black line) of the target (dots) Saturday 10th September.}
  \label{fig:illustration}
\end{figure}

The bootstrap pseudo-realisations can be represented graphically as in Figure \ref{fig:illustration}. This graphic conveys information about the intrinsic variability of the prediction. We can easily detect the instants of day where the heterogeneity of the past situations imply larger uncertainties around the prediction. 

So, this construction is the one used for all the variants proposed to build PIs in the 
sequel by applying them to the constructed set of bootstrap trajectories. 

\subsection{Methods for PI construction} \label{sec:others}

\subsubsection{Two variants from KWF method}

Following \cite{antoniadis2006functional}, the construction of a prediction interval in the stationary case is given by the following procedure:

\begin{enumerate}
\item For $b \in \{ 1, \ldots, B\}$, obtain the pseudo 
      residuals. \\
$\widehat{R}_{n + 1}^{(b)} (t_i) = Z_{n + 1}^{(b)} (t_i) - \widehat{Z}_{n+1}(t_i)$
\item Compute for each $t$ the $\alpha$ and $1-\alpha$ empirical 
      quantiles
$\widehat{R}_{n + 1, \alpha}(t_i)$ and 
      $\widehat{R}_{n+1, 1 - \alpha}(t_i)$
\item For each $t_i$ of the sampling grid, the prediction interval is given by \\ 
$\widehat{L}_{n+1, \alpha} (t_i) 
      = \widehat{R}_{n + 1, \alpha} (t_i)  
      + \widehat{Z}_{n+1} (t_i)  $   \\
 $\widehat{U}_{n+1, \alpha} (t_i)   
      =  \widehat{R}_{n + 1, 1 - \alpha} (t_i)  
      +  \widehat{Z}_{n+1} (t_i) $     
\end{enumerate}

Then in the nonstationary case, the main modifications is to incorporate the 
approximation parts of the segment in a convenient way consistent with the fact that
the detail parts drive the search of similarity. So the procedure becomes:

\begin{enumerate}           
\item For $b \in \{ 1, \ldots, B\}$, obtain the details pseudo 
      residuals of the details and approximations
$\widehat{R}_{n + 1}^{(b)} (t) = 
      \mathcal{D}_{m+1}^{(b)} (t) - \widehat{\mathcal{D}}_{n+1}(t)$ \\
     $\widehat{Q}_{n + 1}^{(b)} (t) = 
      \mathcal{S}_{m+1}^{(b)} (t) - \widehat{\mathcal{S}}_{n+1}(t)$ 
\item Compute for each $t$ the
 $\alpha$ and $1-\alpha$ empirical 
      quantiles for detail residuals $\widehat{R}_{n + 1, \alpha}(t_i)$ and 
      $\widehat{R}_{n+1, 1 - \alpha}(t_i)$
      and select the corresponding approximation parts
      $\widehat{Q}_{n+1, \alpha} (t) $  
      $\widehat{Q}_{n+1, 1 - \alpha} (t)$
\item For each $t_i$ of the sampling grid, the prediction interval is given by       \\
$\widehat{L}_{n+1, \alpha} (t_i)
      =  \widehat{Q}_{n + 1, \alpha} (t_i)  
      +  \widehat{R}_{n + 1, \alpha} (t_i)     
      +  \widehat{Z}_{n+1} (t_i) $ and
 $\widehat{U}_{n+1, \alpha} (t_i)  
      =  \widehat{Q}_{n + 1, 1 - \alpha} (t_i)  
      +  \widehat{R}_{n + 1, 1 - \alpha} (t_i)  
      +  \widehat{Z}_{n+1} (t_i)  $ 
\end{enumerate}

Let us remark that this solution allows to preserve a length of the prediction interval depending
on the specific time instant of the sampling grid, because the connection between approximation 
and details is preserved. But this solution leads to underestimate the variability attributable to 
the approximation part. At the contrary, if we disconnect the two parts by computing
$\widehat{Q}_{n+1, \alpha} (t) $  and $\widehat{Q}_{n+1, 1 - \alpha} (t)$ as the
 $\alpha$ and $1-\alpha$ empirical quantiles for approximation residuals, we obtain larger intervals
 and the price to pay is the homogeneity across time of the length of the prediction interval.
 This was considered in \cite{antoniadis2014prevision}. As this variant produces prediction intervals that are not necessary symmetrical we note it as NS-KWF.

Another possibility considered in \cite{poggi1994prevision} is to construct a symmetric interval of length depending on the standard deviations of the bootstrap residuals and on some theoretical quantile depending
on  $1 - \alpha$. Typically  this is of the form $\widehat{Z_{n+1}} (t_i) \pm 1.96 \hat{\sigma}_{t_i}$ for the symmetric Gaussian quantile at level 0.95 with $\hat{\sigma}_{t_i}$ the empirical standard deviation of $\widehat{Z}^{(b)}_{t_i}$. This variant will be noted as S-KWF.

\subsubsection{Nearest Path heuristic}

Let us present the procedure to obtain simultaneous PI construction using Nearest Path heuristic (NP) proposed by \cite{Staszewska2011}. The underlying idea is to peal out the extremal trajectories from the bunch of bootstrap pseudo-realisations.
We will note this variant NP.

\begin{enumerate}
 \item Obtain a path forecast $\widehat{Z}_{n+1}(t)$ and $B$ bootstrapped forecasts.
 \item Search for extreme paths: at each time point $t_i$, identify 
       the lowest and largest forecast as well as the paths they belong to.
 \item Identify the extremest extreme path using some distance (the original proposition uses the euclidean distance).
 \item Eliminate the extremest extreme path.
 \item If less than $\alpha \times B$ paths were already removed
       go back to 2, otherwise stop and return the envelope of the remaining
       $(1 - \alpha) \times B$ paths as the PI
\end{enumerate}

\subsubsection{Control of the family-wise error ($k$-FWE)}

\cite{Wolf2013} argued that the forecaster may accept that a few time points may not be covered by the simultaneous horizon PI. With this in mind, they construct a joint probability region that controls the family-wise error up to $k$ points. In practice the value of $k$ is decided by the forecaster.
Let us present the procedure to obtain simultaneous PI construction using $k$-FWE.
  
\begin{enumerate}
 \item Obtain a path forecast $\widehat{Z}_{n+1}(t)$ and $B$ bootstrapped forecasts.
 \item Compute the standardized bootstrap residuals 
       $\hat{\mathbf{s}}^\ast_{b} \in \mathbb{R}^H$ with $b=1,\ldots, B$.
 \item Obtain $k\text{-max}_b^\ast$, the $k$ largest value of $|\hat{\mathbf{s}}^\ast_b|$.
 \item Calculate the $1-\alpha$ quantile of $k\text{-max}_b^\ast$. Call it $d\text{-max}$.
 \item The PI for the forecast path is
  \[ \widehat{Z}_{n+1}(t_i) \pm d\text{-max} \cdot \hat{\mathbf{\sigma}}_{t_i} \]
\end{enumerate}

\section{Numerical experiences}  

We describe in this section the main results obtained through the numerical experiences we performed for the construction of prediction intervals for the forecast of the French load curve. 

Applying the methodology described in Section \ref{sec:kwf} we obtain for a whole test year the one-day ahead prediction of the national demand where each daily prediction contains 48 time points. We use the wavelet known as \textit{Symmlet 6} to represent the load curves. The parameter $h_n$ is calibrated through by minimizing a the square loss of the recent past predictions. With the obtained bandwidth, we recuperate the probability weight vector $w_{m, n}$ that we use to produce the 100 bootstrap samples. Then we use each of the methods presented in Section \ref{sec:others} to compute the prediction intervals at three level of confidence 0.80, 0.90 and 0.95. We let the variant $k$-FWE set $k=2$ points outside the construction of the prediction intervals.

There is no consensus on how to validate forecasts. Several indicators have been proposed, see \cite{gneiting2014probabilistic} for a recent review. We will concentrate in two simple indicators that are useful for our practical application : the empirical coverage of the intervals and their amplitude. The seasonal nature of the electricity data and the existence of different day types introduces more complexity to the analysis. 

The KWF prediction method allows also to use a visual check of the variability accounted on the predictor. In fact as we shown in Figure \ref{fig:illustration}, the bunch of curves that enter with a strictly positive weight $w_{m, n}$ allows one to inspect this variability.

\subsection{Global performance}

Tables \ref{tab:amplitude} and \ref{tab:coverage} contain the mean amplitude and mean empirical coverage through the whole year. First, We can see how the amplitudes obtained from the two KWF variants are lower than the k-FWE and NP variants, but this is obtained with an unsatisfactory coverage. 
Second, as it is expected the mean amplitude grows with the confidence level. However, there is a particular case for the NP variant: when it we passes from 80\% to 90\% the variant produces a slight enlargement of the interval that seems to be to small with respect to the increments observed on the other variants. This fact should be compared with the empirical coverage of this variant (88\%) for both of the cited confidence levels.

\begin{table} \centering
\begin{tabular}{cccc} \toprule
Method & 80\% & 90\% & 95\% \\ \midrule
k-FWE  & 3452 & 3837 & 4966 \\
NP     & 4006 & 4038 & 4690 \\
S-KWF  & 2294 & 2620 & 3580 \\
NS-KWF & 2119 & 2377 & 3147 \\ \bottomrule
\end{tabular}
\caption{Mean amplitude (in Mwh)} \label{tab:amplitude}
\end{table}

\begin{table} \centering
\begin{tabular}{cccc} \toprule
Method & 80\% & 90\% & 95\% \\ \midrule
k-FWE  & 87   & 92   & 95 \\
NP     & 88   & 88   & 89 \\
S-KWF  & 73   & 82   & 88 \\
NS-KWF & 67   & 77   & 82 \\ \bottomrule
\end{tabular}
\caption{Mean coverage (in percentage)} \label{tab:coverage}
\end{table}

\begin{figure}
\centering
\includegraphics[width = .8\columnwidth]{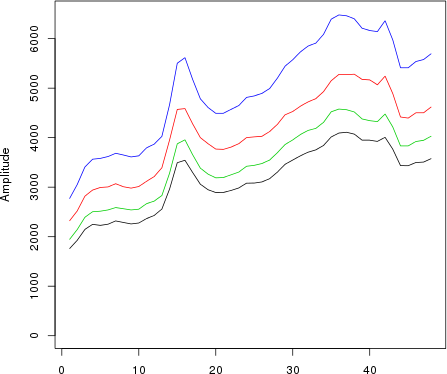}
\caption{Mean amplitude (right) by hour for the variants: k-FWE (blue),
NP (red), S-KWF (green) and NS-KWF (black) using 95\% of confidence level.}
\label{fig:amplitudebyhour}
\end{figure}

\begin{figure}
\centering
\includegraphics[width = .8\columnwidth]{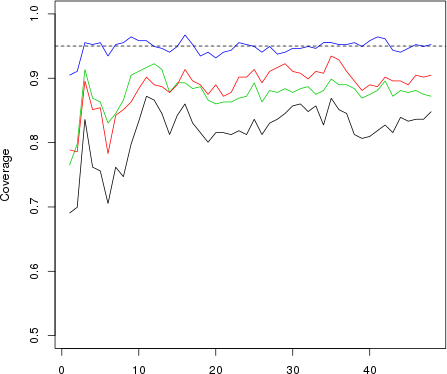}
\caption{Mean coverage by hour for the variants: k-FWE (blue),
NP (red), S-KWF (green) and NS-KWF (black) using 95\% of confidence level.}
\label{fig:coverbyhour}
\end{figure}

In Figures \ref{fig:amplitudebyhour} and \ref{fig:coverbyhour} we present the mean coverage and mean amplitude by hour only for the confidence level of 95\%.
The results remain unchanged, the smaller amplitudes of the variants are obtained for unsatisfactory coverage levels. We
can also notice how the amplitudes are affected by two factors. On one side, the amplitudes tend to grow with the prediction horizon. On the other side, the heterogeneity of the variability of the load curve modulates the first factor: for example enlarging the amplitude during the peak of the morning or reducing it during the lower demand of the night. 

In conclusion, the bootstraping procedure defined on Section \ref{sec:bootKWF} succeeds in capturing the more or less heterogeneity of the situations. 
Despite this fact, the simple use of the KWF variants (symmetrical or non symmetrical) are not sufficient to construct prediction intervals that correctly cover the target. Fortunately, the variants proposed on the literature, in particular $k$-FWE, allow us to construct prediction intervals that correctly cover the path forecast.

\subsection{Some selected cases}

\begin{figure*}
 \centering
 \includegraphics[width=.8\textwidth]{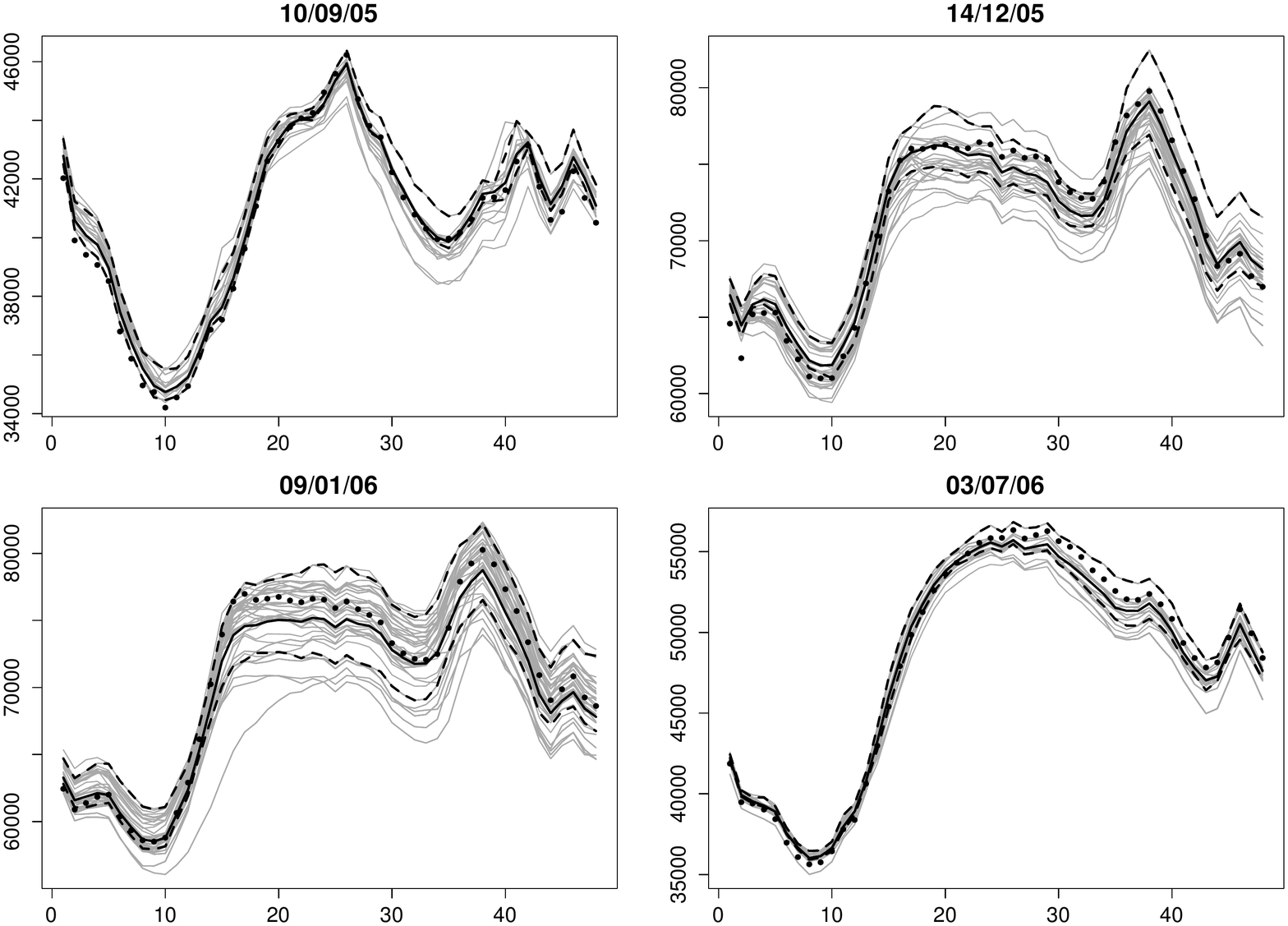}
 \caption{Four selected cases: prediction (black line), boostrap replicates (gray lines) and target (dots).}
 \label{fig:4cases}
\end{figure*}

We perform now a deeper analysis on some selected cases. The four cases we examine are plotted in Figure \ref{fig:4cases}. The two cases on the main diagonal concern narrow prediction intervals with punctual predictions that are close to the targets. The two remaining cases contain larger prediction intervals and larger prediction errors. For each case we plot the punctual prediction, the bootstrap replicates an the target.

Even if we can distinguish two situations where the prediction errors are quite different, it is remarkable that the shape of the predicted load curves is globally well described by the predictions. We recall that no information about the season was used nor any meteorological variable. For the first element, the shape of the curves carries enough information as we shown in Figure \ref{fig:weigths}. Concerning the second element, we should separate the meteorological information already coded in the load curve (the actual situation) and the situation that corresponds to the future. This latter situation is not include in our model. By this way we can explain the two behaviors that we observe in the selected cases. The load curves on the diagonal of Figure \ref{fig:4cases} corresponds to days in July and September which are known to be quite stable on the meteorological conditions. The remaining load curves belongs to the months of December and January where the meteorological conditions can be quite heterogeneous.

\section{Towards the construction of a confidence tube} 

If we place out center of interest on the load curve instead of some of the points
of the curve, we would like to construct the prediction intervals in order to warranty a coverage of the whole prediction path. 
In  Table \ref{tab:curvewise} we report the empirical coverage calculated curve-wise for the confidence level of 95\%, that is we consider that a curve is covered if all of the $48 - k$ points of one curve are covered by their respective prediction intervals. Notice that the forecaster may deem this criterion as acceptable for some small $k$. 
All the variants fail to cover the curve at the announced confidence levels. 
The best performances are obtained by the $k$-FWE variant, which was calibrated to allow up to $k = 2$ points uncovered during the construction of the prediction intervals. It seems that this larger flexibility enables the variant to outperform over the others even if the global result for the curve-wise coverage are not satisfactory.

\begin{table}
\centering
\begin{tabular}{lcccc} \toprule
Variant & k=3  & k=2 & k=1 & k=0\% \\ \midrule
k-FWE   & 84   & 81   & 75 & 70   \\
NP      & 63   & 55   & 48 & 41   \\
S-KWF   & 58   & 53   & 48 & 41   \\
NS-KWF  & 45   & 37   & 31 & 21   \\ \bottomrule
\end{tabular}
\caption{Curve-wise empirical coverage}
\label{tab:curvewise}
\end{table}

Even if this can be only sketched, the study for obtaining a simultaneous PI for $\lbrace Z_{n+1}(t) , t \in [T, T+H]\rbrace$, i.e. a prediction tube for the function-valued forecast is of interest. 
The construction of the prediction tube is a very much difficult problem and fewer references exist. In the particular case where the coordinates of a wavelet expansion of the target function are known to be independent and normally distributed, \cite{Genovese2005} construct a confidence set for $f\in L_2$ a function regression by the coverage the wavelet coefficients $\mu\in \ell_2$ of $f$.
In principle, if we can translate a confidence set for the wavelet coefficients (i.e. a set on $\mathbb{R}^N$) to a confidence set for the function (i.e. a set on $L_2$) a confidence tube can be obtained. 

It seems that a similar problem has been treated from the point of view of hypothesis testing. For example, one can use a multiple hypothesis testing problem to decide which are the non null wavelet coefficients from a noisy signal (see \cite[page 100]{nason2010}).
A similar case is the recent paper of \cite{shen2002}
who use a wavelet representation to capture thee spatial dependence of two signals with the aim of detecting differences by hypothesis testing. The authors propose to increase the power of hypothesis test through the reduction of the number of multiple comparison. If we disposed with a mechanism that allow us to obtain confidence intervals from multiple hypothesis testing we could use these approaches to set up a confidence tube for the prediction.

\section*{Acknowledgments}
We thanks Dr. Stazseska-Bystrova for sharing the computer code that computes the NP heuristic.

\bibliographystyle{plain}
\bibliography{predintervals}


\end{document}